\documentclass[notitlepage,prd,amsfonts,noshowpacs,nofootinbib]{revtex4}
\usepackage{amsmath,graphicx}
\usepackage{color}
\input{colordvi.tex}

\begin{document}
\title{Halo/Galaxy Bispectrum with Equilateral-type Primordial Trispectrum}
\author{Shuntaro Mizuno$^{1}$, and Shuichiro Yokoyama$^{2}$}
\affiliation{$^1$Waseda Institute for Advanced Study, Waseda University, 1-6-1 Nishi-Waseda, Shinjuku, 
	Tokyo 169-8050, Japan,\\
$^2$Department of Physics, Rikkyo University,
3-34-1 Nishi-Ikebukuro, Toshima, Tokyo 171-8501, Japan\\
}

\date{\today}
\begin{abstract}
We investigate the effect of equilateral-type primordial trispectrum on the halo/galaxy bispectrum.
We consider three types of equilateral primordial trispectra which are generated by quartic operators naturally appeared 
in the effective field theory of inflation and  can be characterized by three non-linearity parameters,
$g_{\rm NL} ^{\dot{\sigma}^4}$, $g_{\rm NL} ^{\dot{\sigma}^2 (\partial \sigma)^2}$, and
$g_{\rm NL} ^{(\partial \sigma)^4}$.
Recently, constraints on these parameters have been investigated from CMB observations by using WMAP9 data.
In order to consider the halo/galaxy bispectrum with the equilateral-type primordial trispectra,
we adopt the integrated Perturbation Theory (iPT) in which the effects of primordial
non-Gaussianity are wholly encapsulated in the linear primordial polyspectrum for the  evaluation of the biased polyspectrum.
We show the shapes of the halo/galaxy bispectrum with the equilateral-type primordial trispectra,
and find that the primordial trispectrum characterized by  $g_{\rm NL} ^{\dot{\sigma}^4}$ provides the same scale-dependence as
the gravity-induced halo/galaxy bispectrum.
Hence, it would be difficult to obtain the constraint on $g_{\rm NL} ^{\dot{\sigma}^4}$
from the observations of the halo/galaxy bispectrum.
On the other hand, the primordial trispectra characterized by  $g_{\rm NL} ^{\dot{\sigma}^2 (\partial \sigma)^2}$ and 
$g_{\rm NL} ^{(\partial \sigma)^4}$  provide the common scale-dependence which is different from that of the gravity-induced halo/galaxy bispectrum
on large scales. Hence future observations of halo/galaxy bispectrum would give constraints on the non-linearity parameters,  $g_{\rm NL} ^{\dot{\sigma}^2 (\partial \sigma)^2}$ and 
$g_{\rm NL} ^{(\partial \sigma)^4}$ independently from CMB observations and it is expected that these constraints can be comparable 
to ones obtained by CMB.

\end{abstract}

\maketitle
%%%%%%%%%%%%%%%%%%%%%%%%%%%%%%%%%%%%%%%%%%%%%%%%%%%%
\section{Introduction}
\label{sec:Introduction}
%%%%%%%%%%%%%%%%%%%%%%%%%%%%%%%%%%%%%%%%%%%%%%%%%%%%

The primordial non-Gaussianity provide crucial information on the interaction structure  of inflation (for a review, see \cite{Bartolo:2004if}).
At present, a most stringent constraint on primordial non-Gaussianity is provided by Planck collaboration \cite{Ade:2015ava} and it
implies 
no evidence of non-Gaussianity. Although 
the resultant constraint has almost approached the observational limit predicted by ideal observations,
it is still rather weak
from a particle physics point of view. Therefore, it would be very interesting to try further constraining the non-Gaussianity
based on the information other than CMB.

For this purpose, it has been recently noticed that large-scale halo/galaxy distributions
provide  a distinct information on the primordial non-Gaussianity. Especially, in the presence of local-type primordial non-Gaussianity, 
it has been shown that the halo/galaxy power spectrum is enhanced on large scales (so-called scale-dependent bias), which is helpful to impose the constraint
on the primordial non-Gaussianity (e.g., \cite{Dalal:2007cu,Slosar:2008hx,Matarrese:2008nc}). 
Although the current constraints derived from the scale-dependent bias is still weaker than the one from CMB \cite{Giannantonio:2013uqa}, 
from the future observational projects such as DES \cite{Abbott:2005bi},  BigBoss \cite{Schlegel et al.(2011)},  
LSST \cite{LSST Science Collaboration et al.(2009)},  EUCLID \cite{2011arXiv1110.3193L} and HSC/PFS (Sumire) \cite{2014PASJ...66R...1T}, 
it is expected that we can get the constraint $\Delta f_{\rm NL} ^{\rm local} \sim \mathcal{O} (0.1)$ 
 \cite{Yamauchi:2014ioa}. 
 
 The influence of scale-dependent bias sourced by the primordial non-Gaussianity appears not only in the halo/galaxy 
power spectrum but also in the halo/galaxy bispectrum and other polyspectra. Although it is well known that the late-time nonlinear
gravitational evolution also gives the non-Gaussianity, if the amplitude of primordial non-Gaussianity is sufficiently large,
the halo/galaxy bispectrum sourced by the primordial non-Gaussianity 
has a different scale dependence from the non-linear gravitational evolution and it
can dominate
on large scales \cite{Sefusatti:2007ih,Jeong:2009vd,Sefusatti:2009qh,Baldauf:2010vn,Nishimichi:2009fs,Yokoyama:2013mta,Tasinato:2013vna,Gil-Marin:2014pva,Tellarini:2015faa}.
Especially, when we consider the higher order local-type primordial non-Gaussianity,
by combining the analysis of the halo/galaxy power spectrum with the bispectrum
it is expected that we could get much tighter constraint on the primordial non-Gaussianity.
Another important fact with the halo/galaxy bispectrum is that the amplitude of the contribution sourced by the equilateral-type primordial bispectrum  
is also shown to be enhanced on large scales \cite{Sefusatti:2007ih,Sefusatti:2009qh,Yokoyama:2013mta},
which does not give an enhancement in the halo/galaxy power spectrum.

Regardless of these works, compared with the analysis of CMB, the one of LSS has not covered another important class of
primordial non-Gaussianity, that is, the trispectra generated in theoretical models which produce the equilateral-type bispectrum,
which we call equilateral-type trispectra from now on. This is because the shapes of primordial trispectra of this class strongly depend
on the theoretical models and  they are generically much more complicated than those of the local-type trispectra.
Recently, however, 
Ref. \cite{Smith:2015uia} has investigated an optimal analysis of the such kind of equilateral-type trispectra by making use of CMB observations (for the earlier works to obtain the constraints on the equilateral-type trispectra based on CMB observations, see
Refs.~\cite{Regan:2010cn,Mizuno:2010by,Fergusson:2010gn,Izumi:2011di,Regan:2013jua}).
For the analysis they introduce three new non-linearity parameters, $g_{\rm NL} ^{\dot{\sigma}^4}$, $g_{\rm NL} ^{\dot{\sigma}^2 (\partial \sigma)^2}$, and
$g_{\rm NL} ^{(\partial \sigma)^4}$, which respectively represent the amplitudes of the primordial trispectra that correspond to quartic operators
of the form $\dot{\sigma}^4$, $\dot{\sigma}^2 (\partial \sigma)^2$, and  $(\partial \sigma)^4$ in the effective field theory of inflation (we will show the detailed forms of these trispectra later
in section \ref{sec:HGspectra_with_equil_tri}). 
The reason that only these three trispectra have been considered
is that their forms are relatively simple and they have natural theoretical origin in the sense that they are shown to be generated by
general  $k$-inflation \cite{Huang:2006eha,Arroja:2009pd,Chen:2009bc} and  the effective field theory of inflation 
\cite{Senatore:2010jy,Senatore:2010wk}. 

Following Ref.~\cite{Smith:2015uia}, in this paper, we investigate the effect of these three equilateral-type primordial trispectra on the halo/galaxy bispectrum
and see if we can get  constraints on these trispectra from the future LSS observations independently from those from CMB.
For this purpose,
we adopt the integrated Perturbation Theory (iPT) \cite{Matsubara:2011ck} which enables us to connect the halo/galaxy
clustering with the initial matter density field and incorporate the non-local biasing effect in a straightforward manner
\cite{Yokoyama:2013mta,Matsubara:2012nc,Matsubara:2013ofa,Yokoyama:2012az,Sato:2013qfa}. 
Furthermore, it is worth mentioning that in iPT, we do not rely on the approximations like the peak-background split and
the peak formalism.

This paper is organized as follows. In Sec.~\ref{secHG:spectra_with_primordialNG}, we begin by presenting a general formula
for the halo/galaxy bispectrum in the presence of the primordial bispectrum and trispectrum in terms of iPT. 
In Sec.~\ref{sec:spectra_with_equil_bi},  we show that while the effect of the equilateral-type primordial bispectrum does not appear in the halo/galaxy power spectrum,
it appears in the halo/galaxy bispectrum. For the analysis, we estimate the amplitude of each contribution
based on the the equilateral configuration where the signal becomes maximum. Then, we investigate the effect of the equilateral-type trispectra mentioned above
on the halo/galaxy bispectrum and show that two of them,  $T_\Phi ^ {\dot{\sigma}^2 (\partial \sigma)^2}$ and  $T_\Phi  ^{(\partial \sigma)^4}$
 can give the dominant contribution on very large scales, while  $T_\Phi ^ {\dot{\sigma}^4}$ gives the same scale-dependence
as the one induced by the nonlinearity of the gravitational evolution in Sec.~\ref{sec:HGspectra_with_equil_tri}.
In the same section, we also consider the shape-dependence of the halo/galaxy bispectrum 
to distinguish the effects by the equilateral-type bispectrum from the equilateral-type trispectra
$T_\Phi ^ {\dot{\sigma}^2 (\partial \sigma)^2}$ and  $T_\Phi  ^{(\partial \sigma)^4}$ which provide the common scale-dependence  on large scales
for the equilateral configuration. Sec.~\ref{sec:summary} is devoted to summary.
In our numerical works, through this paper,  we adopt the best fit cosmological parameters taken from Planck \cite{Planck:2015xua}
unless specifically mentioned.

%%%%%%%%%%%%%%%%%%%%%%%%%%%%%%%%%%%%%%%%%%%%%%%%%%%%
\section{Halo/Galaxy spectra with primordial non-Gaussianity}
\label{secHG:spectra_with_primordialNG}
%%%%%%%%%%%%%%%%%%%%%%%%%%%%%%%%%%%%%%%%%%%%%%%%%%%%

In this section, we briefly review the formula for the power- and bi-spectra of galaxies and halos with primordial non-Gaussianity
based on the integrated perturbation theory (iPT). In Sec.~\ref{subsec:HGspectra_from_iPT}, we first present the general expressions
for the power and bispectrum. We keep the terms giving leading contributions up to the one-loop order in iPT.
We then derive the concrete expressions of the multi-point propagators in the large-scale limit in Sec.~\ref{subsec:multi_point_propagators},
which will be the important building blocks to study the scale-dependent behavior of the power
and bispectrum on large scales.

%%%%%%%%%%%%%%%%%%%%%%%%%%%%%%%%%%%%%%%%%%%%%%%%%%%%
\subsection{Halo/Galaxy Power spectrum and Bispectrum from integrated perturbation theory}
\label{subsec:HGspectra_from_iPT}
%%%%%%%%%%%%%%%%%%%%%%%%%%%%%%%%%%%%%%%%%%%%%%%%%%%%

We begin by defining the power- and bi-spectra of biased objects (halos/galaxies),  $P_X$ and  $B_X$:

\begin{eqnarray}
\langle  \delta_X ({\bf k})   \delta_X ({\bf k}')  )  \rangle
&\equiv&  (2 \pi)^3 \delta^{(3)} ({\bf k} + {\bf k}' )  P_X (k)\,,\\
\langle  \delta_X ({\bf k}_1)   \delta_X ({\bf k}_2)   \delta_X ({\bf k}_3 )  \rangle
&\equiv&  (2 \pi)^3 \delta^{(3)} ({\bf k}_1 + {\bf k}_2 + {\bf k}_3)  B_X  ({\bf k}_1, {\bf k}_2, {\bf k}_3)\,,
\end{eqnarray}
where the quantity $\delta_X$ is a Fourier transform of the number density field of the biased objects.
In iPT, the perturbative expansion of the statistical quantities such as power- and bi-spectra of biased objects 
are composed of the multi-point propagators and the polyspectra of the linear density field $\delta_L$.

The definition of the $(n+1)$-point propagator of the biased objects $\Gamma^{(n)} _X$ is given by~\cite{Matsubara:2011ck}
\begin{equation}
\left \langle \frac{\delta^n \delta_X ({\bf k})}{\delta \delta_{\rm L} ({\bf k}_1) \delta \delta_{\rm L} ({\bf k}_2) \cdots
\delta \delta_{\rm L} (({\bf k}_n) } \right \rangle
= (2 \pi)^{3-3n} \delta( {\bf k}_1 + {\bf k}_2 + \cdots + {\bf k}_n) \Gamma^{(n)} _X ({\bf k}_1, {\bf k}_2, \cdots , {\bf k}_n)\,,
\end{equation}
and it represents the influence on $\delta_X$ due to the infinitesimal variation for the initial density field
$\delta_L$ like non-linear gravitational evolution, non-local bias, redshift space distortion etc.
In Sec.~\ref{subsec:multi_point_propagators}, we will show the concrete expression of the 2- and 3-point propagators in the large-scale limit
which play important roles in this paper.

On the other hand, the power-, bi- and tri-spectra of the linear density field $P_{\rm L}$, $B_{\rm L}$ and $T_{\rm L}$
are defined by
\begin{eqnarray}
\langle \delta_{\rm L} ( {\bf k})  \delta_{\rm L} ( {\bf k}') \rangle &=& (2 \pi)^3 \delta( {\bf k} + {\bf k}' ) P_{\rm L} (k)\,,\nonumber\\ 
\langle \delta_{\rm L} ( {\bf k}_1)  \delta_{\rm L} ( {\bf k}_2)  \delta_{\rm L} ( {\bf k}_3) \rangle &=& (2 \pi)^3 \delta( {\bf k}_1 + {\bf k}_2 + {\bf k}_3 ) 
B_{\rm L} ( {\bf k}_1, {\bf k}_2, {\bf k}_3)\,,\nonumber\\
\langle \delta_{\rm L} ( {\bf k}_1)  \delta_{\rm L} ( {\bf k}_2)  \delta_{\rm L} ( {\bf k}_3)  \delta_{\rm L} ( {\bf k}_4) \rangle &=& 
(2 \pi)^3 \delta( {\bf k}_1 + {\bf k}_2 +  {\bf k}_3  +  {\bf k}_4 ) 
T_{\rm L} ( {\bf k}_1, {\bf k}_2, {\bf k}_3 , {\bf k}_4)\,.
\end{eqnarray}

It is worth mentioning that the linear density field $\delta_{\rm L}$ is related to the primordial curvature perturbation $\Phi$ through the function
$\mathcal{M} (k)$:
\begin{equation}
\delta_L (k) = \mathcal{M} (k) \Phi (k); \;\;\;\; {\mathcal{M}} (k) = \frac23 \frac{D(z)}{D(z_*) (1+z_*)} \frac{k^2 T(k)}{H_0 ^2 \Omega_{m0}}\,,
\label{rel_dell_Phi}
\end{equation}
where $T(k)$, $D(z)$, $H_0$ and $\Omega_{m0}$ are the transfer function, the linear growth factor, the Hubble parameter at present epoch,
and the matter density parameter, respectively. $z_*$ denotes an arbitrary redshift at the matter-dominated era.
For the concrete form of the transfer function and the linear growth factor, we use the ones adopted in \cite{Chongchitnan:2010xz} and
\cite{Linder:2005in}, respectively.  Furthermore, because of the finite resolution of any observation, the density field always requires
the procedure of the smoothing over some length scale $R$. For the smoothing, we use the window function $W (kR)$ which is the spherical
top-hat function of $R$, 
\begin{eqnarray}
W(kR) = 3 \left[\frac{\sin (kR)}{(kR)^3} - \frac{\cos (kR)}{(k R)^2} \right]\,, 
\end{eqnarray}
in Fourier space. It is also useful to define the mass scale $M$
\begin{eqnarray}
M \equiv \frac43 \pi R^3 \rho_m \simeq 1.16 \times 10^{12} \Omega_{m0} \left(\frac{R}{h^{-1} {\rm Mpc}} \right)^3 h^{-1} M_{\odot}\,,
\label{rel_RM}
\end{eqnarray}
which is regarded as the mass of matter enclosed by the top-hat window.

With the relation (\ref{rel_dell_Phi}), the linear power spectrum is expressed 
in terms of that of the primordial curvature perturbation as
\begin{equation}
P_{\rm L} (k) = \mathcal{M} (k) ^2 P_{\Phi} (k)\, ,
\end{equation}
with
\begin{equation}
\langle \Phi ( {\bf k}) \Phi  ( {\bf k}') \rangle = (2 \pi)^3 \delta( {\bf k} + {\bf k}' ) P_{\Phi} (k)\,,
\end{equation}
where we assume the scale-invariant primordial power spectrum, that is, $P_\Phi (k) \propto k^{-3}$, for simplicity\footnote{For the equilateral-type trispectrum, a generalisation to the case of the slightly scale-dependent power spectrum has been discussed in Ref.~\cite{Smith:2015uia}.}.
We can define the variance of density fluctuations smoothed on scale $R$ by
\begin{equation}
\sigma_R^2 \equiv \frac{1}{2 \pi^2}\int_0 ^\infty dk k^2 W(kR)^2 \mathcal{M} (k)^2 P_\Phi \,,
\end{equation}
and we choose the normalization of the primordial power spectrum so that it gives
\begin{equation}
\sigma_8 = \sigma(R= 8 h^{-1} {\rm Mpc}) = 0.815 \,,
\end{equation}
which is the value of $\sigma_8$ reported by Planck collaboration  \cite{Planck:2015xua}.

In terms of the multi-point propagators and the linear polyspectra introduced above,
the power spectrum of the biased objects can be written as 
\begin{equation}
P_X (k) = P_0 + P_{\rm bis} + \cdots\,.
\label{def_P_X}
\end{equation}
with
\begin{eqnarray}
P_0 &=& \left[ \Gamma_X ^{(1)} ( {\bf k}) \right]^2 P_{\rm L} (k)\,,\label{def_P_tree}\\
 P_{\rm bis} &=&  \Gamma_X ^{(1)} ( {\bf k}) \int \frac{d^3 p}{(2 \pi)^3}  \Gamma_X ^{(2)} ( {\bf p},  {\bf k} - {\bf p})
B_{\rm L} (\bf{k}, -\bf{p}, -{\bf k} + {\bf p})\,.
\label{def_P_bis}
\end{eqnarray}
Here we have considered the perturbative expansion up to the one-loop order in iPT\footnote{In iPT,  there is another term at one-loop order which is constructed from two $P_{\rm L}$ 
and two $\Gamma^{(2)} _X$. However, since it was shown in \cite{Yokoyama:2012az} that this term is negligible on large scales, we do not consider this term 
in this paper.}. 
Up to the one-loop order in iPT, the contribution from the primordial trispectrum does not appear.
It appears at the two-loop order. However, as shown later, in case with the equilateral-type non-Gaussianity,
the one-loop order contribution given by Eq.~(\ref{def_P_bis}), which is induced by the primordial bispectrum,
is not so significant and it is expected that two-loop order contribution related with the primordial trispectrum would be much suppressed.
Hence, here, for the halo/galaxy power spectrum we neglect the contribution from the equilateral-type primordial trispectrum.

Similarly, the bispectrum of the biased objects can be written as
\begin{equation}
 B_X ({\bf k}_1, {\bf k}_2, {\bf k}_3) = B _{\rm grav} + B _{\rm bis} +  B _{\rm tris} + \cdots\,,
\label{def_B_X}
\end{equation}
with
\begin{eqnarray}
 B _{\rm grav}  &=& \left[\Gamma ^{(1)} _X ( {\bf k}_1) \Gamma ^{(1)} _X ( {\bf k}_2) \Gamma ^{(2)} _X ( -{\bf k}_1,  -{\bf k}_2)
P_{\rm L} (k_1) P_{\rm L} (k_2) + 2\; {\rm perms.}\right]\,,\label{def_B_tree}\\
 B _{\rm bis}  &=& \Gamma ^{(1)}_X ( {\bf k}_1)  \Gamma ^{(1)}_X ( {\bf k}_2)  \Gamma ^{(1)}_X ( {\bf k}_3) 
B_{\rm L} ( {\bf k}_1, {\bf k}_2, {\bf k}_3)\,, \label{def_B_bis}\\ 
 B _{\rm tris} &=& \frac12 
 \Gamma ^{(1)}_X ( {\bf k}_1)  \Gamma ^{(1)}_X ( {\bf k}_2) \int \frac{d^3 p}{(2 \pi)^3}  \Gamma ^{(2)}_X ( {\bf p},  {\bf k}_3- {\bf p} )
T_{\rm L} ( {\bf k}_1, {\bf k}_2, {\bf p} , {\bf k}_3 - {\bf p} )+ 2\; {\rm perms.}\,.
\label{def_B_tris}
\end{eqnarray}
Again, we have considered the perturbative expansion up to the one-loop order in iPT\footnote{In iPT, there are other five terms at one-loop order denoted by $B^{\rm loop,1} _{\rm grav}$,
 $B^{\rm loop,2} _{\rm grav}$, $B^{\rm loop,1} _{\rm bis}$, $B^{\rm loop,2} _{\rm bis}$, $B^{\rm loop,3} _{\rm bis}$ in \cite{Yokoyama:2013mta}.
However, since it was shown in the paper that  all of these terms are negligible on large scales for the case with the equilateral-type primordial bispectrum,
we do not consider these terms in this paper. }  and we find that for the halo/galaxy bispectrum the contribution from the primordial trispectrum appears at the one-loop order.

\begin{figure}[h!]
		\centering
\begin{minipage}{.75\linewidth}
		\centering
		\includegraphics[width=\linewidth]{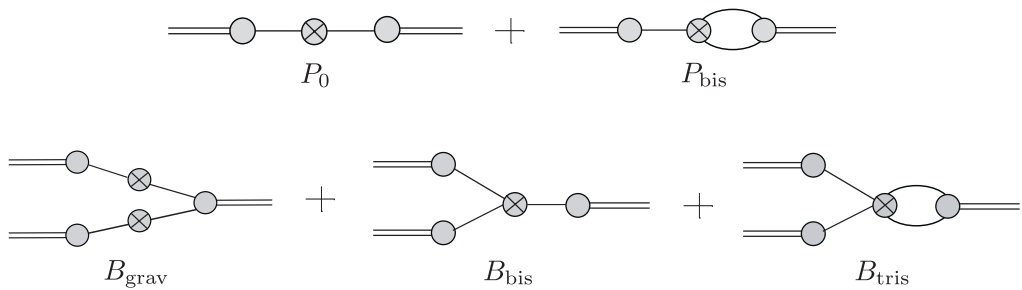}
\end{minipage}
		\caption{Diagrammatic representation of each term in Eqs.~(\ref{def_P_X}) (upper) and (\ref{def_B_X}) (lower).}
		\label{fig:diagram}
	\end{figure}

In Fig.~\ref{fig:diagram}, diagrammatic representation of each term in Eqs.~(\ref{def_P_X}) and (\ref{def_B_X}) is shown.
A double solid line connected with a grey circle indicate the multi-point propagator of biased objects $\Gamma^{(n)} _X$ while
a crossed circle glued to multiple single solid lines indicate the correlator of the initial linear density field.

%>>>>>>>>>>>>>>>>>>>>>>>>>>>>>>>FIGURE<<<<<<<<<<<<<<<<<<<<<<<<<<<<<<<<<<%
%>>>>>>>>>>>>>>>>>>>>>>>>>>>>>>>>>><<<<<<<<<<<<<<<<<<<<<<<<<<<<<<<<<<<<<%

%%%%%%%%%%%%%%%%%%%%%%%%%%%%%%%%%%%%%%%%%%%%%%%%%%%%
\subsection{Multi-point propagators in the large-scale limit}
\label{subsec:multi_point_propagators}
%%%%%%%%%%%%%%%%%%%%%%%%%%%%%%%%%%%%%%%%%%%%%%%%%%%%

The multi-point propagator $\Gamma^{(n)} _X$ is defined as a fully non-perturbative quantity and it is difficult to evaluate
it rigorously. But we know that  the halo/galaxy polyspectra are generically 
dominated by the non-linearity of the gravitational evolution on small scales
and large scales are the only window where the effect 
of the primordial non-Gaussianity can be significant. In such large-scale limit where the scale of interest $\sim 1/k_i$ is 
much larger than the typical scale of the formation of the collapsed object $\sim 1/p$,
the perturbative treatment works well and the multi-point propagators can be simplified as 
\begin{eqnarray}
 \Gamma ^{(1)}_X ( {\bf k})  &\simeq& 1 + c_1 ^{\rm L} (k)\,,\nonumber\\
 \Gamma ^{(2)}_X ( {\bf k}_1,  {\bf k}_2  ) &\simeq& F_2   ( {\bf k}_1,  {\bf k}_2  ) + \left(1+\frac{ {\bf k}_1 \cdot  {\bf k}_2}{k_2 ^2} \right) c_1 ^{\rm L} ( {\bf k}_1)
+  \left(1+\frac{ {\bf k}_1 \cdot  {\bf k}_2}{k_1 ^2} \right) c_1 ^{\rm L} ( {\bf k}_2) +  c_2 ^{\rm L} ( {\bf k}_1,  {\bf k}_2)\,,
\label{expressions_mpp_lsl}
\end{eqnarray}
where $F_2$ is the second-order kernel of standard perturbation theory which is given by 
\begin{eqnarray}
F_2 ( {\bf k}_1,  {\bf k}_2) = \frac{10}{7} + \left( \frac{k_2}{k_1} + \frac{k_1}{k_2}\right) \frac{ {\bf k}_1 \cdot  {\bf k}_2}{k_1 k_2}
+ \frac47 \left( \frac{  {\bf k}_1 \cdot  {\bf k}_2}{k_1 k_2}\right)^2\,.
\end{eqnarray}
Due to the symmetric property of $F_2$, we have
\begin{eqnarray}
 \Gamma ^{(2)}_X ( -{\bf p},  {\bf p}  )  \simeq  c_2 ^{\rm L} ( -{\bf p},  {\bf p})\,.
\end{eqnarray}
In Eq.~(\ref{expressions_mpp_lsl}), $c_n ^{\rm L}$ is a renormalized bias function defined in Lagrangian space, given by
\begin{eqnarray}
c_n ^{\rm L} ( {\bf k}_1,  {\bf k}_2, \cdots,  {\bf k}_n) = (2 \pi)^{3n}
\int \frac{d^3 k'}{(2 \pi)^3} \left \langle \frac{\delta^n \delta^{\rm L} _X ({\bf k}')}{\delta \delta_{\rm L} ({\bf k}_1) \delta \delta_{\rm L} ({\bf k}_2) \cdots
\delta \delta_{\rm L} (({\bf k}_n) } \right \rangle\,,
\end{eqnarray}
where $\delta^{\rm L} _X $ is the number density field of biased objects in Lagrangian space.

For a simple model of non-local halo bias proposed by Ref.~\cite{Matsubara:2011ck,Matsubara:2013ofa}, 
the renormalized bias function for halos with mass $M$ is given by
\begin{eqnarray}
c_n ^{\rm L} ( {\bf k}_1,  {\bf k}_2, \cdots,  {\bf k}_n) = \frac{A_n (M)}{\delta^n _c} W(k_1;M) \cdots  W(k_n;M)
+ \frac{A_{n-1} (M) \sigma^n _M}{\delta^n _c} \frac{d}{ d \ln \sigma_M} 
\left[ \frac{W (k_1; M) \cdots W (k_n; M)}{\sigma^n _M} \right]\,, 
\end{eqnarray}
where $\delta_c (\simeq 1.686)$ is the so-called critical density of the spherical collapse model and 
$\sigma_M$ is the variance of density fluctuations on the mass scale $M$ defined by Eq.~(\ref{rel_RM}).
Here, 
$A_n (M)$ is defined by 
\begin{eqnarray}
A_n (M) \equiv \sum^n _{j=0} \frac{n !}{j !} \delta^j _c b^{\rm L} _j (M)\,,
\end{eqnarray}
where $ b^{\rm L} _j (M)$ is the $j$-th order scale-independent Lagrangian bias parameter which is constructed from the universal
mass function as
\begin{eqnarray}
 b^{\rm L} _j (M) = (- \sigma_M)^{-j} f_{\rm MF} ^{-1}  \frac{d^j}{d \nu^j} (f_{\rm MF} (\nu))\,.
\end{eqnarray}
Throughout the paper, we adopt Sheth-Tormen mass function \cite{Sheth:1999mn} given by
\begin{eqnarray}
f_{\rm ST} (\nu) = A (p) \sqrt{\frac{2}{\pi}} [1 + (q \nu^2)^{-p}] \sqrt{q} \nu e^{-q  \nu^2/2} \,.
\label{sheth_tormen_mf}
\end{eqnarray}
In Eq.~(\ref{sheth_tormen_mf}), $\nu = \delta_c/ \sigma_M$, $p=0.3$, $q = 0.707$ and the normalization factor
$A(p) = [1 + \Gamma (1/2 - p) /(\sqrt{\pi} 2^p) ]^{-1}$.

In the large scale limit where $k_i \to 0$, the window function and its derivative approach
$W(k_i;R) \to 1$ and $dW (k_i;R)/ d \ln \sigma_M \to 0$. Therefore, the renormalized bias function, either the multi-point propagator
does not have significant scale-dependence. Before closing this section, for the later convenience,
it is worth mentioning that in the large-scale limit,  $\mathcal{M} (k)$ appeared in Eq.~(\ref{rel_dell_Phi}) has a scale-dependence
\begin{eqnarray}
\mathcal{M} (k) \propto k^2\,.
\label{asymp_mfunc}
\end{eqnarray}

%%%%%%%%%%%%%%%%%%%%%%%%%%%%%%%%%%%%%%%%%%%%%%%%%%%%
\section{Halo/Glaxy power spectrum and bispectrum with equilateral-type primordial bispectrum}
\label{sec:spectra_with_equil_bi}
%%%%%%%%%%%%%%%%%%%%%%%%%%%%%%%%%%%%%%%%%%%%%%%%%%%%

In this section, based on the simple expressions for the multi-point propagators on large scales which are obtained in the previous section,
we will investigate the effect of equilateral-type primordial bispectrum on the halo/galaxy power spectrum and bispectrum
in Sec~\ref{subsec:HGpower_with_equil_bi} and \ref{subsec:HGbi_with_equil_bi}, in order.

%%%%%%%%%%%%%%%%%%%%%%%%%%%%%%%%%%%%%%%%%%%%%%%%%%%%
\subsection{Halo/Galaxy power spectrum with equilateral-type primordial bispectrum}
\label{subsec:HGpower_with_equil_bi}
%%%%%%%%%%%%%%%%%%%%%%%%%%%%%%%%%%%%%%%%%%%%%%%%%%%%

Among the terms of the halo/galaxy power spectrum in Eq.~(\ref{def_P_X}), $P_0$ generically gives the dominant contribution
on small scales, which means that any type of corrections can be significant only on large scales. Therefore, first let us see the scale-dependence
of  $P_0$ in the large-scale limit. From Eq.~(\ref{def_P_tree}) and making use of the fact that $\Gamma^{(1)} _X ({\bf k})$ has no scale-dependence on large scales, it is estimated as
\begin{eqnarray}
P_0 &\propto& \mathcal{M} (k)^2 P_\Phi \propto k\,.
\label{P_tree_asym}
\end{eqnarray}

On the other hand, in the presence of the primordial bispectrum, the possible correction to  $P_X$
is given by $P_{\rm bis}$ in Eq.~(\ref{def_P_X}). From Eq.~(\ref{def_P_bis}),  in the large-scale limit
$P_{\rm bis}$ can be approximated as
\begin{eqnarray}
P_{\rm bis} &\simeq& \Gamma^{(1)} _{X} ({\bf k}) \int \frac{d^3 p}{(2 \pi)^3} \Gamma^{(2)} _X ({\bf p}, -{\bf p}) B_{\rm L} ({\bf k}, -{\bf p}, {\bf p} )
\nonumber\\
&=& \Gamma^{(1)} _{X} ({\bf k}) \mathcal{M} (k) \int \frac{d^3 p}{(2 \pi)^3} \Gamma^{(2)} _X ({\bf p}, -{\bf p})  \mathcal{M} (p)^2 B_\Phi  ({\bf k}, -{\bf p}, {\bf p} ) \,.
\label{P_bis_interm}
\end{eqnarray} 
Therefore, the scale-dependence of  $P_{\rm bis}$ in the large-scale limit depends on the type of primordial bispectrum. 

It is well known that the effect of the local-type primordial bispectrum whose amplitude is characterized by
the nonlinearity parameter, $f_{\rm NL} ^{\rm local}$, appears in the halo/galaxy power spectrum on large scales.
Actually, by substituting the following shape of the local-type primordial bispectrum \cite{Komatsu:2001rj},
\begin{eqnarray}
B_{\Phi}  ^{\rm local}  ({\bf k}_1,  {\bf k}_2,{\bf k}_3) =  2 f_{\rm NL} ^{\rm local} \left[P_\Phi (k_1) P_\Phi (k_2) +   2\; {\rm perms.}    \right]\,,
\label{shape_localbispectrum}
\end{eqnarray} 
into Eq.~(\ref{P_bis_interm}) and making use of the fact that
neither $\Gamma^{(1)} _X ({\bf k})$  on large scales nor the integral of $p$ in Eq.~(\ref{P_bis_interm}) has no scale-dependence,
we obtain
\begin{eqnarray}
P_{\rm bis}  ^{\rm local} &\propto&  \frac{\mathcal{M} (k)}{k^3}   \propto k^{-1}\,.
\label{P_bis_local_asym}
\end{eqnarray} 
From Eqs.~(\ref{P_tree_asym}) and (\ref{P_bis_local_asym}), we can see that $P_{\rm bis}  ^{\rm local}$ increases
while  $P_0$ decreases as $k$ decreases and we can expect
that  $P_{\rm bis}  ^{\rm local}$ will dominate 
$P_0 $ above some scale, which is called as the scale-dependent bias effect.

However, as we will show, this is not the case with the equilateral-type primordial bispectrum
whose shape is given by \cite{Creminelli:2005hu}
\begin{eqnarray}
B_{\Phi} ^{\rm equil}   ({\bf k}_1,  {\bf k}_2,{\bf k}_3) &=& 6 f_{\rm NL} ^{\rm equil}  
\Biggl[ -\left(P_\Phi (k_1) P_\Phi (k_2) + 2\; {\rm perms.}   \right)\nonumber\\
&&-2 P_\Phi (k_1) ^{2/3} P_\Phi (k_2) ^{2/3}  P_\Phi (k_3) ^{2/3}
+\left( P_\Phi (k_1) ^{1/3}   P_\Phi (k_2) ^{2/3} P_\Phi (k_3)     + 5\; {\rm perms.}     \right) \Biggr]\,.
\label{shape_equilateralbispectrum}
\end{eqnarray} 
Here $ f_{\rm NL} ^{\rm equil} $ is the non-linearity parameter.
Performing the similar procedure as the local-type one, we see that  $B_\Phi  ({\bf k}, -{\bf p}, {\bf p} )  \propto 1/k$
since the terms $\propto 1/k^3$ and $\propto 1/k^2$ in this shape are cancelled because of the high symmetry of this shape. 
Then, we obtain
\begin{eqnarray}
P_{\rm bis} ^{\rm equil} &\propto&  \frac{\mathcal{M} (k)}{k}   \propto k\,.
\label{P_bis_equil_asym}
\end{eqnarray} 
Comparing Eq.~(\ref{P_bis_equil_asym}) with Eq.~(\ref{P_tree_asym}), 
$P_{\rm bis} ^{\rm equil}$ decreases as $k$ decreases with the same scaling as  $P_0$ even in the large-scale limit,
which means that $P_{\rm bis} ^{\rm equil}$ always keeps to be subdominant compared with  $P_0$. Then, we cannot expect that
the effect of equilateral-type primordial bispectrum can be seen through the halo/galaxy power spectrum.

%%%%%%%%%%%%%%%%%%%%%%%%%%%%%%%%%%%%%%%%%%%%%%%%%%%%
\subsection{Halo/Galaxy bispectrum with equilateral-type primordial bispectrum}
\label{subsec:HGbi_with_equil_bi}
%%%%%%%%%%%%%%%%%%%%%%%%%%%%%%%%%%%%%%%%%%%%%%%%%%%%

If there is primordial bispectrum, it naturally affects the halo/galaxy bispectrum.
In  Eq.~(\ref{def_B_X}), this effect is included in $B_{\rm bis} $.
Here, as the shape of the primordial bispectrum, we will consider only the equilateral-type one characterized 
by Eq.~(\ref{shape_equilateralbispectrum})  which was shown to give only a subdominant contribution to the halo/galaxy power spectrum,
\begin{eqnarray}
B_{\rm bis} ^{\rm equil}  &=&  6 f_{\rm NL} ^{\rm equil}  \Gamma ^{(1)}_X ( {\bf k}_1)  \Gamma ^{(1)}_X ( {\bf k}_2)  \Gamma ^{(1)}_X ( {\bf k}_3) 
 \mathcal{M} (k_1)   \mathcal{M} (k_2)  \mathcal{M} (k_3) 
\Biggl[ -\left(P_\Phi (k_1) P_\Phi (k_2) + 2\; {\rm perms.}   \right)\nonumber\\
&&-2 P_\Phi (k_1) ^{2/3} P_\Phi (k_2) ^{2/3}  P_\Phi (k_3) ^{2/3}
+\left( P_\Phi (k_1) ^{1/3}   P_\Phi (k_2) ^{2/3} P_\Phi (k_3)     + 5\; {\rm perms.}     \right) \Biggr]\,.
\end{eqnarray}

On the other hand,  it is well known that although the density fluctuation is Gaussian initially, 
the non-Gaussianity is generated through the non-linearity
of gravitational evolution and this effect is included in $B_{\rm grav} $ in  Eq.~(\ref{def_B_X}).
Since  $B_{\rm grav} $ gives the dominant contribution on small scales, 
we will investigate the amplitude and shape-dependence of  $B_{\rm grav} $ and  $B_{\rm bis} ^{\rm equil}$
in the large-scale limit as in the analysis of the power spectrum.
In Fig.~{\ref{fig:shapeBbisequilBgrav}, we plot $B_{\rm grav} $ and  $B_{\rm bis} ^{\rm equil}$ to show the shape of each contribution in $k$-space.
We fix $k_1 = 0.003 h {\rm Mpc}^{-1}$ and set the redshift and the mass scale of halos to $z=1.0$ and $M=5 \times 10^{13} h^{-1} M_{\odot}$,
respectively. For the information of halo, we use these values throughout this paper. 
We take $f_{\rm NL} ^{\rm equil} = 80$, which is almost the 2-$\sigma$ upper bound obtained by Planck collaboration \cite{Ade:2015ava}.
Notice that from the symmetry and the triangle condition, it is enough to consider only $k_1 \geq k_2 \geq k_3$ and $k_3 \geq k_1 - k_2$.

%>>>>>>>>>>>>>>>>>>>>>>>>>>>>>>>FIGURE<<<<<<<<<<<<<<<<<<<<<<<<<<<<<<<<<<%
	\begin{figure}[h!]
		\centering
\begin{minipage}{.45\linewidth}
		\centering
		\includegraphics[width=\linewidth]{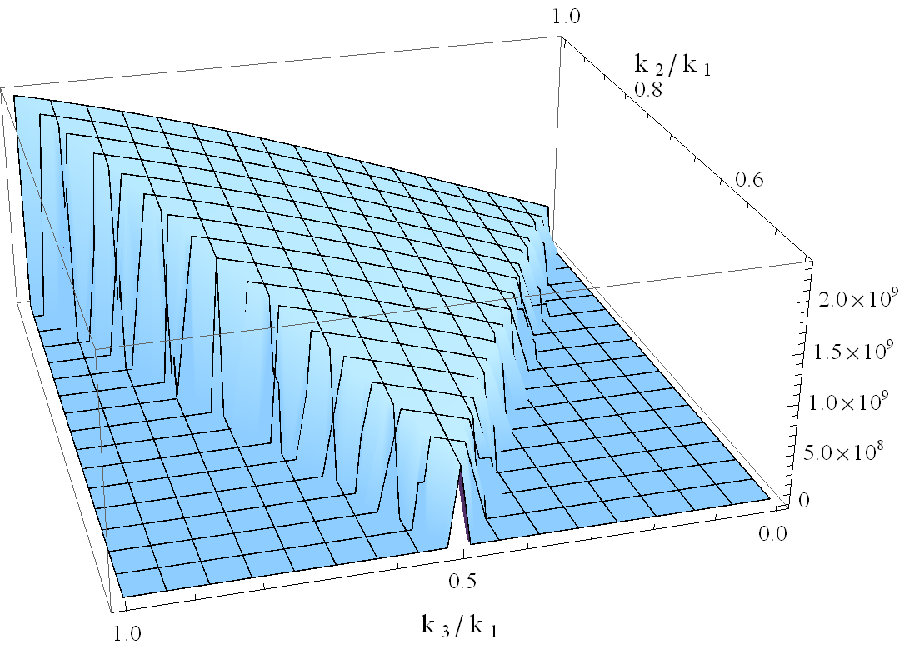}
\end{minipage}
\begin{minipage}{.45\linewidth}
		\centering
		\includegraphics[width=\linewidth]{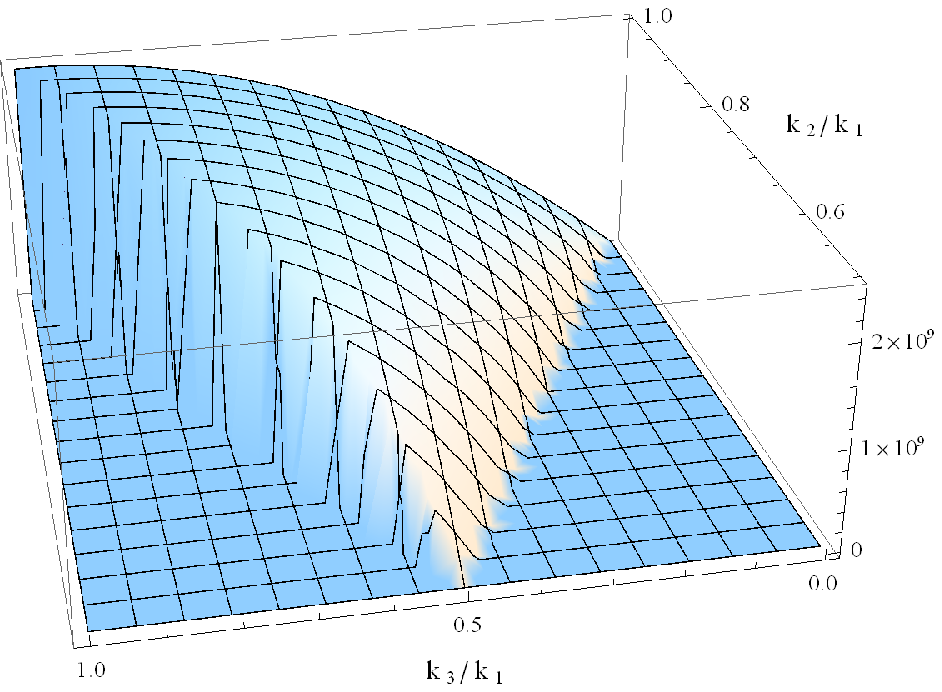}
\end{minipage}
		\caption{The shapes of  $B_{\rm grav} $ (left panel) and  $B_{\rm bis} ^{\rm equil}$ (right panel)
as functions of $k_2/k_1$ and $k_3/k_1$ in momentum space for $k_1 = 0.003  h {\rm Mpc}^{-1}$ .
 We adopt $f_{\rm NL} ^{\rm equil} = 80$.}
		\label{fig:shapeBbisequilBgrav}
	\end{figure}
%>>>>>>>>>>>>>>>>>>>>>>>>>>>>>>>>>><<<<<<<<<<<<<<<<<<<<<<<<<<<<<<<<<<<<<%

From Fig.~\ref{fig:shapeBbisequilBgrav}, we can see that both  $B_{\rm grav} $ and  $B_{\rm bis} ^{\rm equil}$ take the maximum values
at the equilateral configuration ($k_1 = k_2 = k_3$). Therefore, in order to clarify the scale-dependence of their contributions,
we concentrate on the equilateral configuration given by $k \equiv k_1 = k_2 = k_3$.

Then, from Eq.~(\ref{def_B_tree}) and making use of the fact that the multi-point propagators have no scale-dependence on large scales
after fixing the configuration,
the scale-dependence of  $B_{\rm grav} ^{\rm tree}$ is estimated as 
\begin{eqnarray}
B_{\rm grav} &\propto& \mathcal{M} (k)^4 P_\Phi ^2 \propto k^2\,,
\label{B_tree_asym}
\end{eqnarray} 
while from Eq.~(\ref{def_B_bis}) and the similar procedure, the scale-dependence of  $B_{\rm bis} ^{\rm equil}$ is estimated as 
\begin{eqnarray}
B_{\rm bis} ^{\rm equil} &\propto& \mathcal{M} (k)^3 P_\Phi ^2 \propto k^0\,.
\label{B_bis_asym}
\end{eqnarray} 

From Eqs.~(\ref{B_tree_asym}) and (\ref{B_bis_asym}), we can see that $B_{\rm bis} ^{\rm equil}$ keeps to be constant
while  $B_{\rm grav}$ decreases as $k$ decreases and we can expect that  $B_{\rm bis} ^{\rm equil}$ will dominate 
$B_{\rm grav}  $ above some scale. For the quantitative analysis, 
we plot the contributions $B_{\rm grav} $ and $B_{\rm bis} ^{\rm equil}$ which we obtain numerically as functions of the wavenumber $k$ 
in Fig.~\ref{fig:BbisequilvsBgrav}. We can see that for  $f_{\rm NL} ^{\rm equil} = 80$,  $B_{\rm bis} ^{\rm equil}$ dominates $B_{\rm grav} $
at $k \alt 0.003 h {\rm Mpc}^{-1}$.

%>>>>>>>>>>>>>>>>>>>>>>>>>>>>>>>FIGURE<<<<<<<<<<<<<<<<<<<<<<<<<<<<<<<<<<%
	\begin{figure}[h!]
		\centering
\begin{minipage}{.45\linewidth}
		\centering
		\includegraphics[width=\linewidth]{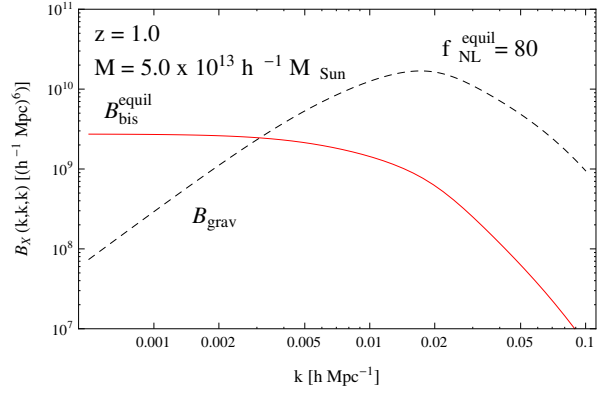}
\end{minipage}
		\caption{ $B_{\rm grav} $ (black dashed line) and  $B_{\rm bis} ^{\rm equil}$ (red thick line)
as a function of $k$. We take the equilateral configuration characterized by  $k = k_1 = k_2 = k_3$ 
and adopt $f_{\rm NL} ^{\rm equil} = 80$.}
		\label{fig:BbisequilvsBgrav}
	\end{figure}
%>>>>>>>>>>>>>>>>>>>>>>>>>>>>>>>>>><<<<<<<<<<<<<<<<<<<<<<<<<<<<<<<<<<<<<%

%%%%%%%%%%%%%%%%%%%%%%%%%%%%%%%%%%%%%%%%%%%%%%%%%%%%
\section{Halo/Glaxy bispectrum with equilateral-type primordial trispectra}
\label{sec:HGspectra_with_equil_tri}
%%%%%%%%%%%%%%%%%%%%%%%%%%%%%%%%%%%%%%%%%%%%%%%%%%%%

In the previous section, we confirm the fact that we could see the effect of  the equilateral-type primordial bispectrum
through the halo/galaxy bispectrum if $f_{\rm NL} ^{\rm equil}$ takes the value of the current $2 \sigma$ upper bound. 

Then, let us focus on the halo/galaxy bispectrum with equilateral-type primordial trispectrum,
which appears at the one-loop order in iPT.
Generally, inflation models that produce equilateral-type primordial bispectrum
also produce primordial $trispectrum$.
  After imposing scale-invariance, the trispectrum is described by a scalar function of five scalar variables,
while the bispectrum is by two scalar variables. Therefore, although the current constraints are still very limited,
the information of the primordial trispectra is helpful to constrain such inflation models.
In this section, we investigate whether we could see the effect of the equilateral-type primordial $trispectra$ through the halo/galaxy bispectrum.

Among the primordial trispectra which can be generated by models producing the equilateral-type bispectrum,  
we concentrate on the following three types of trispectra:
\begin{eqnarray}
T_\Phi ^{\dot{\sigma}^4}  ({\bf k}_1,  {\bf k}_2,{\bf k}_3,{\bf k}_4) &=& \frac{221184}{25}\; g_{\rm NL} ^{\dot{\sigma}^4}\;A_\Phi^3\;S^{\dot{\sigma}^4}
 ({\bf k}_1,  {\bf k}_2,{\bf k}_3,{\bf k}_4)\,,\label{tris_c1}\\
T_\Phi ^{\dot{\sigma}^2 (\partial \sigma)^2}  ({\bf k}_1,  {\bf k}_2,{\bf k}_3,{\bf k}_4)&=& -\frac{27648}{325}\; g_{\rm NL} ^{\dot{\sigma}^2 (\partial \sigma)^2}\;  A_\Phi^3\;
S^{\dot{\sigma}^2 (\partial \sigma)^2} ({\bf k}_1,  {\bf k}_2,{\bf k}_3,{\bf k}_4) \,,\label{tris_c2}\\
T_\Phi ^{(\partial \sigma)^4}  ({\bf k}_1,  {\bf k}_2,{\bf k}_3,{\bf k}_4) &=& \frac{16588}{2575} \;g_{\rm NL} ^{(\partial \sigma)^4} \; A_\Phi^3\; S^{(\partial \sigma)^4}
 ({\bf k}_1,  {\bf k}_2,{\bf k}_3,{\bf k}_4) \,,\label{tris_c3}
\end{eqnarray} 
with
\begin{eqnarray}
S^{\dot{\sigma}^4}  ({\bf k}_1,  {\bf k}_2,{\bf k}_3,{\bf k}_4) 
&=& \frac{1}{\left(\sum_{i=1} ^4 k_i \right)^5 \Pi_{i=1} ^4 k_i}\,,\label{shape_c1}\\
S^{\dot{\sigma}^2 (\partial \sigma)^2}  ({\bf k}_1,  {\bf k}_2,{\bf k}_3,{\bf k}_4)
 &=& \frac{k_1^2 k_2^2 ( {\bf k}_3 \cdot  {\bf k}_4)}{\left(\sum_{i=1} ^4 k_i \right)^3 \Pi_{i=1} ^4 k_i ^3}
\left( 1 + 3 \frac{k_3 + k_4}{   \sum_{i=1} ^4 k_i    } +12 \frac{k_3  k_4}{\left(\sum_{i=1} ^4 k_i \right)^2 }\right)
+5\;\; {\rm perms.}\,,\label{shape_c2}\\
S^{(\partial \sigma)^4}  ({\bf k}_1,  {\bf k}_2,{\bf k}_3,{\bf k}_4) 
&=& \frac{  ({\bf k}_1 \cdot  {\bf k}_2) ({\bf k}_3 \cdot  {\bf k}_4) +  ({\bf k}_1 \cdot  {\bf k}_3) ({\bf k}_2 \cdot  {\bf k}_4) +
 ({\bf k}_1 \cdot  {\bf k}_4) ({\bf k}_2 \cdot  {\bf k}_3) }{ \sum_{i=1} ^4 k_i  \Pi_{i=1} ^4 k_i ^3 }\nonumber\\
&&\times \left(1 + \frac{ \sum_{i<j} k_i k_j}{\left(\sum_{i=1} ^4 k_i \right)^2} + 
3 \frac{  \Pi_{i=1} ^4 k_i }{\left(\sum_{i=1} ^4 k_i \right)^3}   \sum_{i=1} ^4 \frac{1}{k_i} + 
12 \frac{  \Pi_{i=1} ^4 k_i }{   \left(\sum_{i=1} ^4 k_i \right)^4 }\right)\,.
\label{shape_c3}
\end{eqnarray} 
Here, $ g_{\rm NL} ^{\dot{\sigma}^4}$, $g_{\rm NL} ^{\dot{\sigma}^2 (\partial \sigma)^2} $ and $g_{\rm NL} ^{(\partial \sigma)^4}$
are non-linearity parameters which characterize the amplitude of each trispectrum, $ A_\Phi$ is the amplitude of the primordial power spectrum, defined by 
$A_\Phi = k^3 P_\Phi$.  In Eqs.~(\ref{tris_c1}), (\ref{tris_c2}) and (\ref{tris_c3}),
the normalization have been chosen so that they give $(216/25) g_{\rm NL} A_\Phi^3/k^9$ for tetrahedral
4-point configurations with $| {\bf k}_i | = k$ and $  {\bf k}_i \cdot  {\bf k}_j = -k^2/3$ for $i \neq j$. This convention fixes all trispectra
to have the same values on the tetrahedron as the local trispectrum.

Before starting the analysis, we briefly explain the physical motivation for concentrating on the above three trispectra. First, it was shown that
these trispectra are generated by general $k$-inflation models through the contact interaction which is characterized
by a quartic vertex \cite{Huang:2006eha}.  But it turned out that these trispectra are just a part of the full trispectra for this type of inflation models
and they were completed to add another type of trispectra generated through the scalar-exchange interaction which is characterized by two cubic vertices 
\cite{Arroja:2009pd,Chen:2009bc}. From this result, 
it was pointed out that the amplitude of $T_\Phi ^{\dot{\sigma}^4}$ can be large even 
when the equilateral-primordial bispectrum is small by tuning the model parameters. 
This possibility was supplemented by the effective field theory of inflation \cite{Cheung:2007st} to clarify the symmetry
that keeps to give  $T_\Phi ^{\dot{\sigma}^4}$ while protects the generation of cubic terms which are related with the other trispectra.
In this respect, the trispectrum  $T_\Phi ^{\dot{\sigma}^4}$ was regarded as more important than the other trispectra
generated by models producing the equilateral-type primordial bispectrum. Actually, the constraints on this trispectrum imposed by WMAP5
were reported in \cite{Fergusson:2010gn}. 

However, recently, a new possibility that the three trispectra $T_\Phi ^{\dot{\sigma}^4}$, $T_\Phi ^{\dot{\sigma}^2 (\partial \sigma)^2}$
and $T_\Phi ^{(\partial \sigma)^4}$ are equally important  in the context of the effective field theory of $multi$-field inflation  \cite{Smith:2015uia}.
In this set-up,  since we can protect the cubic interactions, the other trispectra generated through the scalar-exchange interaction are suppressed. 
In the same paper, the authors also perform the optimal analysis of the CMB trispectrum and impose the constraints on the 
non-linearity parameters for these three trispectra making use of the fact that the shapes of these trispectra can be written
as factorizable forms, which enables us to reduce the computational cost.  
Following   \cite{Smith:2015uia}, we will concentrate on the case that these three trispectra are equally important, while 
the other trispectra related with the cubic terms are suppressed. 
Although our analysis from now on is completely
phenomenological in the sense that we regard the non-linearity parameters $g_{\rm NL}$ to be free, for those who are interested in
how these trispectra are obtained in concrete models, we show the trispectra
generated by general $k$-inflation models through the contact interaction in Appendix~\ref{sec:equil_tri}.

The effect of the primordial trispectrum on the halo/galaxy bispectrum is given by $B_{\rm tris}$ in Eq.~(\ref{def_B_X}).
From Eq.~(\ref{def_B_tris}),  in the large-scale limit
$B_{\rm tris}$ can be approximated as
\begin{eqnarray}
B_{\rm tris} &\simeq&  \frac12 
 \Gamma ^{(1)}_X ( {\bf k}_1)  \Gamma ^{(1)}_X ( {\bf k}_2) \int \frac{d^3 p}{(2 \pi)^3}  \Gamma ^{(2)}_X ( {\bf p},  - {\bf p} )
T_{\rm L} ( {\bf k}_1, {\bf k}_2, {\bf p} ,  - {\bf p} )+ 2\; {\rm perms.}
\nonumber\\
&=& \frac12  \Gamma ^{(1)}_X ( {\bf k}_1) \mathcal{M} (k_1) \Gamma ^{(1)}_X ( {\bf k}_2)  \mathcal{M} (k_2) \int \frac{d^3 p}{(2 \pi)^3}  \Gamma ^{(2)}_X ( {\bf p},  - {\bf p} )
\mathcal{M} (p)^2 T_{\rm \Phi} ( {\bf k}_1, {\bf k}_2, {\bf p} ,  - {\bf p} )+ 2\; {\rm perms.} \,.
\label{B_tris_interm}
\end{eqnarray} 

Then Substituting Eqs.~(\ref{tris_c1}), (\ref{tris_c2}) and (\ref{tris_c3}) into Eq.~(\ref{B_tris_interm}) gives
\begin{eqnarray}
\frac{B_{\rm tris} ^{\dot{\sigma}^4}}{A_\Phi^3}  &\simeq& \frac{3456}{25}\; g_{\rm NL} ^{\dot{\sigma}^4}
  \Gamma ^{(1)}_X ( {\bf k}_1)  \Gamma ^{(1)}_X ( {\bf k}_2) \frac{\mathcal{M} (k_1)}{k_1} \frac{ \mathcal{M} (k_2)}{k_2} \int \frac{d^3 p}{(2 \pi)^3}  \Gamma ^{(2)}_X ( {\bf p},  - {\bf p} ) \frac{\mathcal{M} (p)^2}{p^7} + 2\; {\rm perms.}\,,\label{Btrisc1}\\
\frac{B_{\rm tris} ^{\dot{\sigma}^2 (\partial \sigma)^2}}{A_\Phi^3}  &\simeq&  -\frac{1728}{325} g_{\rm NL} ^{\dot{\sigma}^2 (\partial \sigma)^2}  
 \Gamma ^{(1)}_X ( {\bf k}_1)  \Gamma ^{(1)}_X ( {\bf k}_2)  \frac{\mathcal{M} (k_1)}{k_1 ^3} \frac{ \mathcal{M} (k_2)}{k_2 ^3} ({\bf k}_1 \cdot {\bf k}_2)
 \int \frac{d^3 p}{(2 \pi)^3}  \Gamma ^{(2)}_X ( {\bf p},  - {\bf p} ) \frac{\mathcal{M} (p)^2}{p^5} + 2\; {\rm perms.}\,,\label{Btrisc2}\\
\frac{B_{\rm tris} ^{(\partial \sigma)^4} }{A_\Phi^3}  &\simeq& -\frac{4147}{2060} g_{\rm NL} ^{(\partial \sigma)^4} 
 \Gamma ^{(1)}_X ( {\bf k}_1) \Gamma ^{(1)}_X ( {\bf k}_2) \frac{\mathcal{M} (k_1)}{k_1 ^3}  \frac{ \mathcal{M} (k_2)}{k_2 ^3}
 \int \frac{d^3 p}{(2 \pi)^3}  \Gamma ^{(2)}_X ( {\bf p},  - {\bf p} ) \frac{\mathcal{M} (p)^2}{p^5}\nonumber\\
&& \times \Biggl[ ({\bf k}_1 \cdot {\bf k}_2) + 2 \left(\frac{ {\bf p}}{p} \cdot {\bf k}_1 \right) \left(\frac{ {\bf p}}{p} \cdot {\bf k}_2 \right)\Biggr]
 + 2\; {\rm perms.}\nonumber\\
&=& -\frac{4147}{1236} g_{\rm NL} ^{(\partial \sigma)^4}
 \Gamma ^{(1)}_X ( {\bf k}_1)  \Gamma ^{(1)}_X ( {\bf k}_2)  \frac{\mathcal{M} (k_1)}{k_1 ^3} \frac{ \mathcal{M} (k_2)}{k_2 ^3} ({\bf k}_1 \cdot {\bf k}_2)
 \int \frac{d^3 p}{(2 \pi)^3}  \Gamma ^{(2)}_X ( {\bf p},  - {\bf p} ) \frac{\mathcal{M} (p)^2}{p^5} + 2\; {\rm perms.} \,,\label{Btrisc3}
\end{eqnarray} 
where in the last line of Eq.~(\ref{Btrisc3}), we have used the relation about the angular part of the integration of $p$
\begin{eqnarray}
\int d\Omega_p  \left(\frac{ {\bf p}}{p} \cdot {\bf k}_1 \right) \left(\frac{ {\bf p}}{p} \cdot {\bf k}_2 \right) = \frac{4 \pi}{3}  ({\bf k}_1 \cdot {\bf k}_2) 
=\frac{ ({\bf k}_1 \cdot {\bf k}_2) }{3} \int d\Omega_p\,.
\end{eqnarray}
From Eqs.~(\ref{Btrisc1}), (\ref{Btrisc2}) and (\ref{Btrisc3}), we can easily see that although we have started with three equilateral-types of the primordial trispectra,
in the large-scale limit, $B_{\rm tris} ^{\dot{\sigma}^2 (\partial \sigma)^2}$ and $B_{\rm tris} ^{(\partial \sigma)^4}$ become  degenerate
and we get only two types of shapes in the halo/galaxy bispectrum. This is caused by the fact the primordial trispectrum 
$T_\Phi ^{\dot{\sigma}^2 (\partial \sigma)^2}$ is very strongly correlated with   $T_\Phi ^{(\partial \sigma)^4}$ and from this reason,
only two of the three trispectra,  $T_\Phi ^{\dot{\sigma}^4}$ and  $T_\Phi ^{(\partial \sigma)^4}$ were used 
as the basis of the optimal analysis of the CMB trispectrum \cite{Ade:2015ava,Smith:2015uia}. 
From this reason, we will concentrate on the two 
equilateral-type primordial trispectra  $T_\Phi ^{\dot{\sigma}^4}$ and  $T_\Phi ^{(\partial \sigma)^4}$ 
where the constraints from CMB have been obtained.
Notice that although we do not mention the effect of $T_{\Phi} ^{\dot{\sigma}^2 (\partial \sigma)^2}$ from now on,
once we can constrain the effect of  $T_{\Phi}   ^{(\partial \sigma)^4}$, it should be constrained by the similar degree. 

 Then, from Eqs.~(\ref{Btrisc1}) and (\ref{Btrisc3}), and 
making use of the fact that neither $\Gamma^{(1)} _X ({\bf k})$  on large scales nor the integral of $p$ in Eq.~(\ref{P_bis_interm}) has no scale-dependence,
we can obtain the following scale-dependence of $B_{\rm tris} ^{\dot{\sigma}^4} $
and $B_{\rm tris} ^{(\partial \sigma)^4} $: 
\begin{eqnarray}
B_{\rm tris} ^{\dot{\sigma}^4} &\propto&  \frac{\mathcal{M} (k)^2}{k^2}  \propto k^2\,, \label{Btrisc1_equil}\\
B_{\rm tris} ^{(\partial \sigma)^4}  &\propto& 
 \frac{\mathcal{M} (k)^2}{k ^4}  
 \propto k^0\,.\label{Btrisc2c3_equil}
\end{eqnarray}
As shown in the previous section, $B_{\rm grav}$ has the scale-dependence which is proportional to $k^2$ in large scales.
Comparing the above scale-dependent behaviors of $B_{\rm tris} ^{\dot{\sigma}^4}$ and $B_{\rm tris} ^{(\partial \sigma)^4} $ with that of $B_{\rm grav}$,  we can expect that 
$B_{\rm tris} ^{(\partial \sigma)^4}$ will dominate $B_{\rm grav}$ above some scale, while it is difficult to find $B_{\rm tris} ^{\dot{\sigma}^4}$
which has the same scale-dependence as $B_{\rm grav}$. Thus, hereinafter we focus on the halo/galaxy bispectrum with the primordial trispectrum $B_{\rm tris} ^{(\partial \sigma)^4}$.

In Fig.~{\ref{fig:shapeBtris}, we plot
$B_{\rm tris} ^{(\partial \sigma)^4}$  to show not only the $k$-dependence with $k = k_1 = k_2 = k_3$, but also the shape of each contribution in $k$-space.
We fix $k_1 = 0.003 h {\rm Mpc}^{-1}$  and take 
$g_{\rm NL} ^{(\partial \sigma)^4}= 2.0 \times 10^7$ so that it gives almost the same amplitude as  $B_{\rm bis}^{\rm equil}$ with  $f_{\rm NL} ^{\rm equil} = 80$.

%>>>>>>>>>>>>>>>>>>>>>>>>>>>>>>>FIGURE<<<<<<<<<<<<<<<<<<<<<<<<<<<<<<<<<<%
	\begin{figure}[h!]
\begin{minipage}{.45\linewidth}
		\centering
		\includegraphics[width=\linewidth]{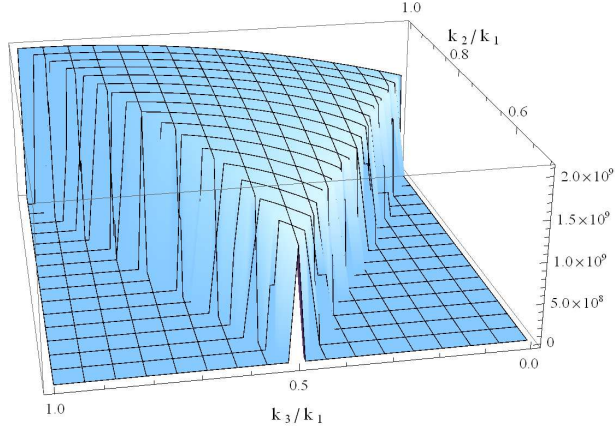}
\end{minipage}
		\caption{The shape of  
		$  B_{\rm tris} ^{(\partial \sigma)^4} $
as a function of $k_2/k_1$ and $k_3/k_1$ in momentum space for  $k_1 = 0.003  h {\rm Mpc}^{-1}$ .
 We adopt  $g_{\rm NL} ^{(\partial \sigma)^4}= 2.0 \times 10^7$.}
		\label{fig:shapeBtris}
	\end{figure}
%>>>>>>>>>>>>>>>>>>>>>>>>>>>>>>>>>><<<<<<<<<<<<<<<<<<<<<<<<<<<<<<<<<<<<<%

From Fig.~\ref{fig:shapeBtris}, we can see that
  $B_{\rm tris} ^{(\partial \sigma)^4}$
takes the maximum value
at the equilateral configuration ($k_1 = k_2 = k_3$)  as is the case in $B_{\rm grav}$ and $B_{\rm bis}^{\rm equil}$. 
Therefore, first
we concentrate on the equilateral configuration given by $k \equiv k_1 = k_2 = k_3$.
For the quantitative analysis,  we plot the contributions $B_{\rm grav} $ and $B_{\rm tris} ^{(\partial \sigma)^4}$ 
which we obtain numerically as functions of the wavenumber $k$ 
in Fig.~\ref{fig:BtrisequilvsBgrav}.
We can see that 
for  $g_{\rm NL} ^{(\partial \sigma)^4}= 2.0 \times 10^7$,  $B_{\rm tris} ^{(\partial \sigma)^4}$ dominates   $B_{\rm grav} $
at $k \alt 0.003 h {\rm Mpc}^{-1}$, and if we can observe such large scales, we can detect this, in principle.
 
%>>>>>>>>>>>>>>>>>>>>>>>>>>>>>>>FIGURE<<<<<<<<<<<<<<<<<<<<<<<<<<<<<<<<<<%
	\begin{figure}[h!]
		\centering
\begin{minipage}{.45\linewidth}
		\centering
		\includegraphics[width=\linewidth]{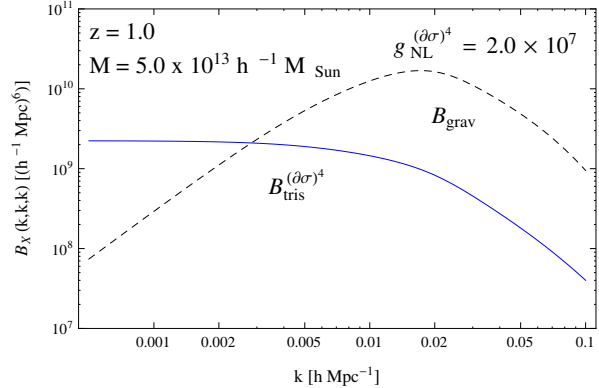}
\end{minipage}
		\caption{ $B_{\rm grav} $ (black dashed line) and $ B_{\rm tris} ^{(\partial \sigma)^4} $ (blue thick line) 
as functions of $k$. We take the equilateral configuration characterized by  $k = k_1 = k_2 = k_3$ 
and adopt $g_{\rm NL} ^{(\partial \sigma)^4}= 2.0 \times 10^7$ 
.}
		\label{fig:BtrisequilvsBgrav}
	\end{figure}
%>>>>>>>>>>>>>>>>>>>>>>>>>>>>>>>>>><<<<<<<<<<<<<<<<<<<<<<<<<<<<<<<<<<<<<%

In the above discussion, we confirm the fact that we could see the effect of  one of the equilateral-type primordial trispectra,
labelled as $T_\Phi ^{(\partial \sigma)^4} $, through the halo/galaxy bispectrum on much larger scales if $g_{\rm NL} ^{(\partial \sigma)^4}$ 
is about $O(10^7)$.  
On the other hand, comparing Figs.~\ref{fig:BbisequilvsBgrav} and \ref{fig:BtrisequilvsBgrav},
we see that both $B_{\rm bis} ^{\rm equil}$ and $B_{\rm tris}  ^{(\partial \sigma)^4}$ have the same scale-dependence $\propto k^0$, which means that
it is difficult to distinguish these two effects as long as we only consider the equilateral configuration. 

However,  as Figs.~\ref{fig:shapeBbisequilBgrav}
and \ref{fig:shapeBtris}, the two shapes of  $B_{\rm bis} ^{\rm equil}$ and $B_{\rm tris}  ^{(\partial \sigma)^4}$ in Fourier space
are different. Especially, 
the amplitude of $B_{\rm tris} ^{(\partial \sigma)^4}$ does not decrease so much at $k_1 = k_2 = 2 k_3$, so-called folded configuration
and this feature is very different from that of $B_{\rm bis}^{\rm equil}$.
Hence, we expect that in principle  by considering a different configuration it would be possible to distinguish the contributions from   $B_{\rm bis} ^{\rm equil}$ and $B_{\rm tris}  ^{(\partial \sigma)^4}$
in the halo/galaxy bispectrum.
  For this purpose, we introduce the isosceles
configuration given by $k\equiv k_1=k_2= \alpha k_3$ and characterized by a parameter $\alpha$. 
The parameter $\alpha$ can take $\alpha \geq 1/2$ and $\alpha = 1$ corresponds to the equilateral configuration.

%>>>>>>>>>>>>>>>>>>>>>>>>>>>>>>>FIGURE<<<<<<<<<<<<<<<<<<<<<<<<<<<<<<<<<<%
	\begin{figure}[h!]
		\centering
\begin{minipage}{.45\linewidth}
		\centering
		\includegraphics[width=\linewidth]{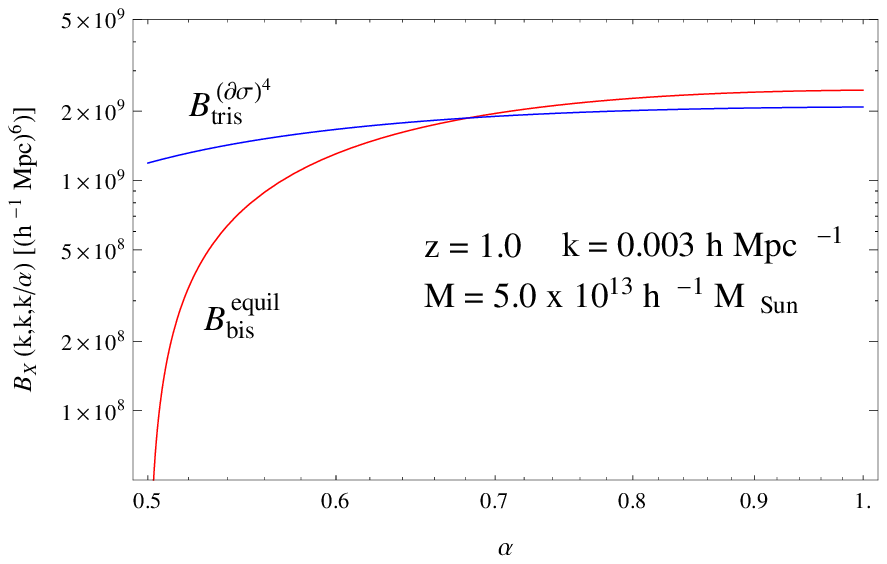}
\end{minipage}
\begin{minipage}{.45\linewidth}
		\centering
		\includegraphics[width=\linewidth]{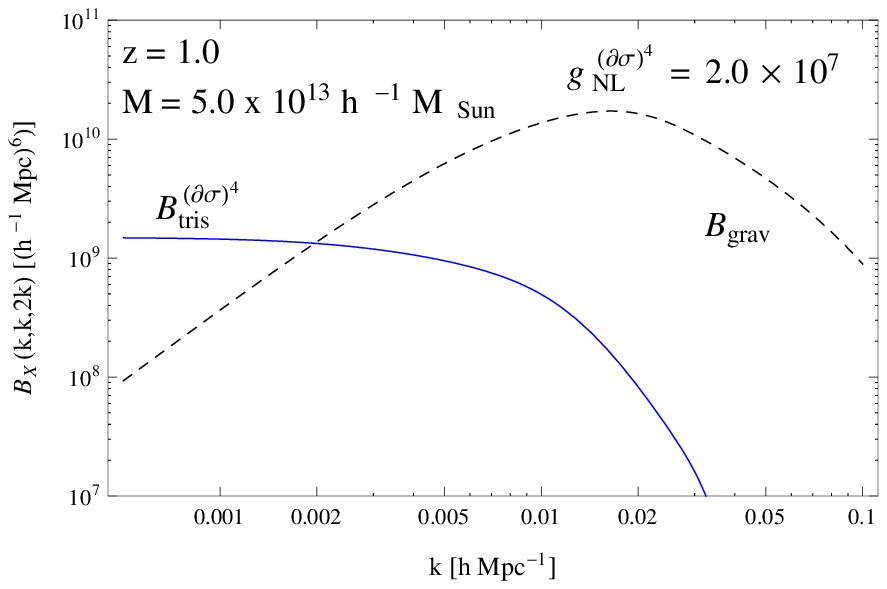}
\end{minipage}
		\caption{(Left panel) $ B_{\rm bis} ^{\rm equil} $ (red line) and  $B_{\rm tris} ^{(\partial \sigma)^4}$ (blue line) 
as functions of $\alpha$ which characterizes  the isosceles
configuration given by $k\equiv k_1=k_2= \alpha k_3$. We take $k=0.003 h {\rm Mpc}^{-1}$ and
adopt $f_{\rm NL} ^{\rm equil} =80$ and 
$g_{\rm NL} ^{(\partial \sigma)^4}= 2.0 \times 10^7$. (Right panel)  $B_{\rm grav} $ (black dashed line) and  $B_{\rm tris} ^{(\partial \sigma)^4}$ (blue thick line) 
and as functions of $k$. We take the folded configuration characterized by  $k = k_1 = k_2 = k_3/2$ 
and adopt $g_{\rm NL} ^{(\partial \sigma)^4}= 2.0 \times 10^7$. }
		\label{fig:alpha_dependence}
	\end{figure}
%>>>>>>>>>>>>>>>>>>>>>>>>>>>>>>>>>><<<<<<<<<<<<<<<<<<<<<<<<<<<<<<<<<<<<<%

In the left panel of Fig.~\ref{fig:alpha_dependence},  we plot the contributions  $B_{\rm bis} ^{\rm equil}$
and  $B_{\rm tris} ^{(\partial \sigma)^4}$ as functions of the parameter $\alpha$.
We can see that while  $B_{\rm bis} ^{\rm equil}$ is comparable to  $B_{\rm tris} ^{(\partial \sigma)^4}$ at the equilateral configuration ($\alpha=1$),
it falls to zero very quickly at the folded configuration ($\alpha = 1/2$).
Therefore, even if there is primordial bispectrum whose effect gives the same scale-dependence of halo/galaxy bispectrum ($\propto k^0$)
at the equilateral configuration, we can eliminate this effect by considering the folded configuration.
Therefore, if  $B_{\rm tris} ^{(\partial \sigma)^4}$ can dominate $B_{\rm grav}$ on large scales also at the folded configuration,
we can see the effect of this type of primordial trispectra.
In the right panel of Fig.~\ref{fig:alpha_dependence}, we confirm that this actually happens as in the case of the equilateral configuration.
Therefore, by considering both equilateral and folded configurations, we can see the effect of the primordial trispectrum  $T_\Phi ^{(\partial \sigma)^4} $
through the halo/galaxy bispectrum if its amplitude is sufficiently large.

%%%%%%%%%%%%%%%%%%%%%%%%%%%%%%%%%%%%%%%%%%%%%%%%%%%%
\section{Summary and Discussions}
\label{sec:summary}
%%%%%%%%%%%%%%%%%%%%%%%%%%%%%%%%%%%%%%%%%%%%%%%%%%%%

The information contained in the primordial non-Gaussianity will contribute to a huge advance in our understanding
of the physics of inflation. Although recent CMB observation by the Planck satellite has reported a very stringent
constraints on the primordial non-Gaussianity \cite{Ade:2015ava}, it would be very interesting  to try further constraining the amplitude
of non-Gaussianity based on the information other than CMB. For this purpose, recently, the fact that the large scale halo/galaxy distributions
are affected by the primordial non-Gaussianity through the scale-dependent bias has been paid much attention. Although there have been many
important works on investigating the effect of primordial non-Gaussianity on the scale-dependence of halo/galaxy distributions,
the most works have been restricted to the primordial bispectrum and local-type trispectrum. This is because the shapes of the equilateral-type
primordial trispectra strongly depend on theoretical models and also because their forms are generically much more complicated than
those of the local-type trispectrum. Regardless of this, since this class of primordial trispectrum possess more information of the 
interaction structure of inflation, it would be worth trying to constrain this class of trispectrum, too. In this line, recently, based on the optimal analysis of 
the CMB, constraints on the amplitudes of the three equilateral-type trispectra 
$T_\Phi ^ {\dot{\sigma}^4}$, $T_\Phi ^ {\dot{\sigma}^2 (\partial \sigma)^2}$ and $T_\Phi  ^{(\partial \sigma)^4}$  
have been obtained in Ref.~\cite{Smith:2015uia}.
These trispectra are considered not just because their forms are relatively simple, but have natural theoretical origin in the sense that they are shown to be appeared 
related with  general  $k$-inflation \cite{Huang:2006eha,Arroja:2009pd,Chen:2009bc}  and effective field theory of inflation 
\cite{Senatore:2010jy,Senatore:2010wk}.

In this paper, we have investigated the effect of these three important equilateral-type primordial trispectra on the scale-dependence
of large scale halo/galaxy distributions. For this purpose, we have adopted the iPT formalism by which we can calculate
the non-local biasing effect in the presence of any types of primordial non-Gaussianity systematically. 
Since it is not necessary for us to rely on the approximations like the peak background split and the peak formalism in iPT,
the formulation for the large scale halo/galaxy distributions based on iPT can provide more general results than the formalisms
mentioned above. 

Before considering the effect of equilateral-type primordial trispectrum,
we have demonstrated that it is necessary to consider the halo/galaxy bispectrum to see the scale-dependent behavior of halo/galaxy
distributions sourced by the equilateral-type primordial bispectrum. This is completely different from the cases with the local-type primordial non-Gaussianity
where there is an enhancement of the halo/galaxy power spectrum on large scales. We have shown that this difference comes from the fact that
the shape of equilateral-type bispectrum has higher symmetry than the one of local-type bispectrum, which cancels the component enhanced on large scales
in the halo/galaxy power spectrum. Since it is expected that a similar statement holds for the equilateral-type primordial trispectrum,
we have investigated the effect of such trispectrum on the halo/galaxy bispectrum.

For the analysis of the scale-dependence of the halo/galaxy bispectrum in the presence of equilateral-type primordial trispectrum, 
although we had started with three primordial trispectra
$T_\Phi ^ {\dot{\sigma}^4}$, $T_\Phi ^ {\dot{\sigma}^2 (\partial \sigma)^2}$ and $T_\Phi  ^{(\partial \sigma)^4}$ ,
we have found that the large scale behaviors of $B_{\rm tris} ^ {\dot{\sigma}^2 (\partial \sigma)^2}$ and $B_{\rm tris}  ^{(\partial \sigma)^4}$,
the contributions sourced by $T_\Phi ^ {\dot{\sigma}^2 (\partial \sigma)^2}$ and $T_\Phi  ^{(\partial \sigma)^4}$, respectively, 
become degenerate and we have got only two independent shapes. This is related with the fact that
the primordial trispectrum  $T_\Phi ^ {\dot{\sigma}^2 (\partial \sigma)^2}$ is very strongly correlated with  $T_\Phi  ^{(\partial \sigma)^4}$
and only two trispectra, $T_\Phi ^ {\dot{\sigma}^4}$ and  $T_\Phi  ^{(\partial \sigma)^4}$ had been used as the basis of the optimal analysis
of the CMB trispectrum  \cite{Ade:2015ava,Smith:2015uia}. 
We have found that $B_{\rm tris} ^ {\dot{\sigma}^2 (\partial \sigma)^2}$ and $B_{\rm tris}  ^{(\partial \sigma)^4}$ are enhanced on large scales
and dominate  $B_{\rm grav} $, the contribution induced by the nonlinearity of the gravitational evolution, 
on very large scales. On the other hand,  we have shown that  $B_{\rm tris} ^ {\dot{\sigma}^4}$, the contribution sourced by $T_\Phi ^ {\dot{\sigma}^4}$ 
has the same scale-dependence as  $B_{\rm grav} $ and it cannot be expected that  we can find $B_{\rm tris} ^ {\dot{\sigma}^4}$.
Actually, for  $g_{\rm NL} ^{(\partial \sigma)^4} = 2.0 \times 10^7$ with which  $B_{\rm tris}  ^{(\partial \sigma)^4}$ gives almost the same
amplitude as $B_{\rm bis} ^{\rm equil}$ with $f_{\rm NL} ^{\rm equil} =80$, almost the 2-$\sigma$ upper bound obtained by Planck collaboration \cite{Ade:2015ava},
$B_{\rm tris}  ^{(\partial \sigma)^4}$ would dominate the halo/galaxy bispectrum on large scales. 
Setting the typical redshift and the mass
of the halos in surveys to be $z=1.0$ and $M=5 \times 10^{13} h^{-1} M_{\odot}$, respectively,  $B_{\rm tris}  ^{(\partial \sigma)^4}$
with  $g_{\rm NL}  ^{(\partial \sigma)^4} = 2.0 \times 10^7$ will dominate  $B_{\rm grav} $ at $k \alt 0.003 h {\rm Mpc}^{-1}$. 
So far, we have estimated the scale-dependence of the halo/galaxy bispectrum with an equilateral configuration where the amplitudes
of the contributions take the maximum values. But we have seen that $B_{\rm bis} ^{\rm equil}$,  $B _{\rm tirs} ^  {\dot{\sigma}^2 (\partial \sigma)^2}$
and  $B _{\rm tirs} ^{(\partial \sigma)^4}$
provide the same scale-dependence on large scales. 
In order to pick up only the information of 
 $T_\Phi  ^{(\partial \sigma)^4}$, we have shown that the folded configuration where  $B_{\rm bis} ^{\rm equil}$ falls to zero very quickly
is helpful. 
 
 In summary, in this paper, it has been shown that we can constrain the non-linear parameters  $g_{\rm NL}  ^{(\partial \sigma)^4}$ and 
$g_{\rm NL} ^ {\dot{\sigma}^2 (\partial \sigma)^2}$ by the future LSS observations independently from those from CMB and we can use this
at least as cross check of the CMB results. Next natural question is whether  the constraints based on the future LSS observations can be 
more stringent than the ones from CMB. Actually, according to  \cite{Smith:2015uia}, the 2-$\sigma$ upper bound obtained by WMAP9 data
is $0.19 \times 10^6$. Given the fact that it is expected that the future LSS observations can constrain $f_{\rm NL} ^{\rm equil} \sim \mathcal{O}(10)$ 
\cite{Sefusatti:2007ih}, and a simple extrapolation provides   $g_{\rm NL}  ^{(\partial \sigma)^4}$, $g_{\rm NL} ^ {\dot{\sigma}^2 (\partial \sigma)^2}$ 
$\sim \mathcal{O}(10^6)$, which is almost the same order as the ones obtained by current CMB observations. However, as we have shown that 
$B_{\rm tris}  ^{(\partial \sigma)^4} $  and $B_{\rm tris}  ^ {\dot{\sigma}^2 (\partial \sigma)^2} $ have signal for wider regions in $k$-space
than $B_{\rm bis} ^{\rm equil}$, which may provide more stringent constraints on   
$g_{\rm NL}  ^{(\partial \sigma)^4}$, $g_{\rm NL} ^ {\dot{\sigma}^2 (\partial \sigma)^2}$. We leave the discussion on the detailed analysis
to estimate the forecast on $g_{\rm NL}  ^{(\partial \sigma)^4}$, $g_{\rm NL} ^ {\dot{\sigma}^2 (\partial \sigma)^2}$  to future work.

Finally, as is mentioned above, we have concentrated on three equilateral-type primordial trispectra whose amplitudes are constrained by CMB observations
and theoretical origin is very clear. But there are still many interesting primordial trispectra generated by theoretical models
which produce the equilateral-type bispectrum  \cite{Gao:2009gd,Mizuno:2009cv,Mizuno:2009mv,Chen:2009zp,Huang:2010ab,Izumi:2010wm,Bartolo:2010di,Izumi:2010yn,Gao:2010xk,Creminelli:2010qf,Renaux-Petel:2013wya,Renaux-Petel:2013ppa,Fasiello:2013dla,Bartolo:2013eka,Arroja:2013dya}. Although constraints are not obtained for these trispectra
even by CMB observations, it might be interesting to consider the possibility to constrain these primordial trispectra based on the 
large scale halo/galaxy distributions.

%%%%%%%%%%%%%%%%%%%%%%%%%%%%%%%%%%%%%%%%%%%%%%%%%%%%%%%%%%%%%%%%%%%%%%%%%%%%%%%%%%%%%%%%%%%%%%%%%%%%%%%%%%%%%%%%%%%%%%%%%%%
\begin{acknowledgments}
S.M.  is supported by JSPS Grant-in-Aid for Research Activity Start-up No. 26887042.
The authors thank T. Matsubara and A. Taruya for useful discussions.

\end{acknowledgments}

\appendix

%%%%%%%%%%%%%%%%%%%%%%%%%%%%%%%%%%%%%%%%%%%%%%%%%%%%
\section{Equilateral-type primordial trispectrum in general single-field $k$-inflation models}
\label{sec:equil_tri}
%%%%%%%%%%%%%%%%%%%%%%%%%%%%%%%%%%%%%%%%%%%%%%%%%%%%

Here, we briefly summarize the primordial trispectra
generated by the general single-field $k$-inflation models  \cite{Arroja:2009pd} (see also \cite{,Chen:2009bc}). 
The action of $k$-inflation is given by
\begin{eqnarray}
S = \frac12 \int d^4x \sqrt{-g} \left[R + 2 P(X, \phi) \right]\,.
\label{action_kinf}
\end{eqnarray} 
where $R$ is the Ricci scalar, $\phi$ is the inflaton field, $X \equiv -(1/2) g^{\mu \nu} \partial_\mu \phi \partial_\nu \phi$ is its kinetic term.

We calculate the primordial trispectrum  making use of the so-called ``interaction picture formalism" \cite{Weinberg:2005vy}.
As we mentioned in Sec.~\ref{sec:HGspectra_with_equil_tri}, although there are two types of trispectra which are generated through
the contact interaction  characterized by a quartic vertex and the scalar-exchange interaction  characterized by two cubic vertices,
we concentrate on the former ones.
For this class of models, the fourth-order interaction Hamiltonian of the field perturbation $\sigma \equiv \delta \phi$
in the flat gauge at leading order in the slow-roll expansion  are given by
\begin{eqnarray}
H_I ^{(4)} (\eta) = \int d^3 x \left[\beta_1 { \sigma_I'} ^4 + \beta_2   {\sigma_I'} ^2 (\partial  \sigma_I)^2 + \beta_3 (\partial  \sigma_I)^4  \right]\,.
\label{fourth_order_int_hamiltonian}
\end{eqnarray} 
where the subscript $I$ denotes that the variable is evaluated in the interaction picture, 
the prime denotes derivative with respect to conformal time $\eta$ and coefficients $\beta_1$, $\beta_2$ and $\beta_3$
are given by
\begin{eqnarray}
\beta_1 &=& P_{,XX} \left(1-\frac98 c_s^2 \right) - 2X P_{,XXX} \left(1-\frac34 c_s^2 \right)+ \frac{X^3 c_s^2}{P_{,X}} P^2_{,XXX}-\frac16 X^2 P_{,4X}\,,
\label{def_beta1}\\
\beta_2 &=& -\frac12 P_{,XX} \left(1-\frac32 c_s^2 \right) + \frac12 X c_s^2 P_{,XXX}\,,\label{def_beta2}\\
\beta_3 &=& -\frac{c_s^2}{8} P_{,XX}\,,\label{def_beta3}
\end{eqnarray} 
where $c_s$ is the sound speed given by
\begin{eqnarray}
c_s^2 = \frac{P_{,X}}{P_{,X} + 2 X P_{,XX}}\,.
\label{sound_speed}
\end{eqnarray} 

Based on this interaction Hamiltonian,
we can calculate  the primordial trispectrum of the inflaton field perturbation at horizon crossing as
\begin{eqnarray}
\langle \Omega |  \sigma (0, {\bf k}_1)   \sigma (0, {\bf k}_2)   \sigma (0, {\bf k}_3)   \sigma (0, {\bf k}_4)   | \Omega \rangle
= -i \int_{-\infty} ^{0} d \eta \langle 0|
\left[ \sigma_I  (0, {\bf k}_1)   \sigma_I  (0, {\bf k}_2)  \sigma_I  (0, {\bf k}_3)  \sigma_I  (0, {\bf k}_4), H_I ^{(4)} (\eta)    \right]
| 0\rangle\,,
\label{trispectrum_field}
\end{eqnarray} 
where $| \Omega \rangle$ denotes the vacuum in the interaction picture.

At leading order in slow-roll and in the small sound speed limit,
in order to obtain the primordial trispectrum of the curvature perturbation at some time after horizon crossing,
we can use the linear relation $\Phi = (H/\dot{\phi}) \delta \phi$ because the higher order terms in this relation only generate sub-leading corrections to this
result. Then, we can obtain the following equilateral-type primordial trispectra:
\begin{eqnarray}
&&\langle \Omega |  \Phi (0, {\bf k}_1)   \Phi (0, {\bf k}_2)   \Phi (0, {\bf k}_3)   \Phi (0, {\bf k}_4)   | \Omega \rangle\nonumber\\
&&= - (2 \pi)^3 \delta^{(3)}  ( {\bf k}_1 +   {\bf k}_2 +  {\bf k}_3 +  {\bf k}_4 ) A_\Phi ^3 \frac{X}{c_s P_{,X}}
\left(1152 \beta_1 c_s^3 S^{\dot{\sigma}^4} +  \beta_2 c_s S^{\dot{\sigma}^2 (\partial \sigma)^2} 
+32 \beta_3 c_s^{-1}  S^{(\partial \sigma)^4}\right)\,,
\label{trispectrum_Phi}
\end{eqnarray} 
where $A_\Phi \equiv k^3 P_{\Phi} (k) = H^4/(4 X c_s P_{,X})$ is the amplitude of the primordial power spectrum
and  $S^{\dot{\sigma}^4}$, $S^{\dot{\sigma}^2 (\partial \sigma)^2}$ and $ S^{(\partial \sigma)^4}$ are shape functions
given by Eqs.~(\ref{shape_c1}), (\ref{shape_c2}) and (\ref{shape_c3}), respectively.

By comparing Eq.~(\ref{trispectrum_Phi}) with Eqs.~(\ref{tris_c1}), (\ref{tris_c2}), and (\ref{tris_c3}),
we can express the non-linear parameters $ g_{\rm NL} ^{\dot{\sigma}^4}$, $ g_{\rm NL} ^{\dot{\sigma}^2 (\partial \sigma)^2}$
and $g_{\rm NL} ^{(\partial \sigma)^4}$ in terms of the derivatives of $P$ with respect to $X$. However, since we have considered
general $k$-inflation model so far and kept $P$ to be an arbitrary function of $\phi$ and $X$, it is not easy to see which
trispectrum can give the dominant contribution among the three in Eq.~(\ref{trispectrum_Phi}).
In order to see this, we consider the DBI inflation as a concrete example \cite{Silverstein:2003hf} where 
the functional form of $P(\phi, X)$ is given by
\begin{eqnarray}
P(X,\phi) = -f(\phi)^{-1} \sqrt{1- 2 f(\phi) X} -V(\phi)\,,
\label{action_dbi}
\end{eqnarray} 
where $f(\phi)$ and $V(\phi)$ are functions of $\phi$ determined by string theory configurations, 
the derivatives of $P$ is related with $c_s$ like $c_s = P_{,X}^{-1}$. Then, at leading order in the sound speed,
$\beta_1$, $\beta_2$ and $\beta_3$ are simplified as
\begin{eqnarray}
\beta_1 = \frac{1}{4 c_s^7 X}\,,\,\;\;\;\;\beta_2 = \frac{1}{8 c_s^3 X}\,,\,\;\;\;\;\beta_3 = -\frac{1}{16 c_s X}\,.
\label{beta_dbi}
\end{eqnarray} 
Therefore, from Eqs.~(\ref{trispectrum_Phi}) and (\ref{beta_dbi}), we can see  $T^{\dot{\sigma}^4} _\Phi$
gives the dominant contribution and the other two terms $ T^{\dot{\sigma}^2 (\partial \sigma)^2} _\Phi $ and $T^{(\partial \sigma)^4} _\Phi$
are subdominant unless $1/c_s^2 \sim 1$, 
in which case the trispectrum is only marginally large $\sim \mathcal{O} (1)$.
Although we do not show explicitly, similar things happens  and  the contributions from 
$ T^{\dot{\sigma}^2 (\partial \sigma)^2} _\Phi $ and $T^{(\partial \sigma)^4} _\Phi$ cannot be dominant 
whenever we can expect large non-Gaussian signal \cite{Chen:2009bc} in general single-field $k$-inflation.
However, as we explain in  Sec.~\ref{sec:HGspectra_with_equil_tri}, the result based on the effective theory of multifield inflation  \cite{Senatore:2010wk}
suggests that we can realize the situation where the three trispectra  $T^{\dot{\sigma}^4} _\Phi$,  $ T^{\dot{\sigma}^2 (\partial \sigma)^2} _\Phi $
and  $T^{(\partial \sigma)^4} _\Phi$ give comparable contributions if we consider multi-field extension of the $k$-inflation models.

%%%%%%%%%%%%%%%%%%%%%%%%%%%%%%%%%%%%%%%%%%%%%%%%%%%%%%%%%%%%%%%%%%%%%%%%%%%%%%%%%%%%%%%%%%%%%%%%%%%%%%%%%%%%%%%%%%%%%
%%%%%%%%%%%%%%%%%%%%%%%%%%%%%%%%%%%%%%%%%%%%%%%%%%%%%%%%%%%%%%%%%%%%%%%%%%%%%%%%%%%%%%%%%%%%%%%%%%%%%%%%%%%%%%%%%%%%%%%%%%%
%%%%%%%%%%%%%%%%%%%%%%%%%%%%%%%%%%%%%%%%%%%%%%%%%%%%%%%%%%%%%%%%%%%%%%%%%%%%%%%%%%%%%%%%%%%%%%%%%%%%%%%%%%%%%%%%%%%%%%%%%%%
%%%%%%%%%%%%%%%%%%%%%%%%%%%%%%%%%%%%%%%%%%%%%%%%%%%%%%%%%%%%%%%%%%%%%%%%%%%%%%%%%%%%%%%%%%%%%%%%%%%%%%%%%%%%%%%%%%%%%%%%%%%
%%%%%%%%%%%%%%%%%%%%%%%%%%%%%%%%%%%%%%%%%%%%%%%%%%%%%%%%%%%%%%%%%%%%%%%%%%%%%%%%%%%%%%%%%%%%%%%%%%%%%%%%%%%%%%%%%%%%%%%%%%%
%%%%%%%%%%%%%%%%%%%%%%%%%%%%%%%%%%%%%%%%%%%%%%%%%%%%%%%%%%%%%%%%%%%%%%%%%%%%%%%%%%%%%%%%%%%%%%%%%%%%%%%%%%%%%%%%%%%%%%%%%%%
%%%%%%%%%%%%%%%%%%%%%%%%%%%%%%%%%%%%%%%%%%%%%%%%%%%%%%%%%%%%%%%%%%%%%%%%%%%%%%%%%%%%%%%%%%%%%%%%%%%%%%%%%%%%%%%%%%%%%%%%%%%
%%%%%%%%%%%%%%%%%%%%%%%%%%%%%%%%%%%%%%%%%%%%%%%%%%%%%%%%%%%%%%%%%%%%%%%%%%%%%%%%%%%%%%%%%%%%%%%%%%%%%%%%%%%%%%%%%%%%%%%%%%%
%%%%%%%%%%%%%%%%%%%%%%%%%%%%%%%%%%%%%%%%%%%%%%%%%%%%%%%%%%%%%%%%%%%%%%%%%%%%%%%%%%%%%%%%%%%%%%%%%%%%%%%%%%%%%%%%%%%%%%%%%%%
%%%%%%%%%%%%%%%%%%%%%%%%%%%%%%%%%%%%%%%%%%%%%%%%%%%%%%%%%%%%%%%%%%%%%%%%%%%%%%%%%%%%%%%%%%%%%%%%%%%%%%%%%%%%%%%%%%%%%%%%%%%

\end{document}